\documentclass[11pt]{article} 
\usepackage[mathscr]{eucal}
\usepackage{latexsym}
\usepackage{epsfig}
\usepackage{amsmath}
\usepackage{amssymb}
\usepackage{amscd}
\newcommand{\M}{{m}}

\newcommand{\oG}{\overline \mG}
\newcommand{\mG}{{\mathcal G}}
\newcommand{\oS}{\overline \mS}
\newcommand{\oF}{\overline \mF}
\newcommand{\mS}{\mathcal S}
\newcommand{\mF}{{\mathcal F}}
\newcommand{\N}{{n}}
\newcommand{\R}{{\mathcal R}} 
\newcommand{\rSn}{{\rm S}_\N}
\newcommand{\otG}{\overline {\tG}}
\newcommand{\oSn}{\overline \rSn}

\newcommand{\hS}{\Omega}

\newcommand{\tG}{\widetilde {\rm G}}
\newcommand{\sd}{\ltimes}
\newcommand{\Q}{\mathcal Q}
\newcommand{\tQ}{\widetilde \Q}
\newcommand{\Hil}{\mathbb H} 
\newcommand{\re}{\mathbb R}
\newcommand{\Z}{\mathbb Z}

\newcommand{\hN}{\widehat P}
\newcommand{\hmF}{\widehat \mF} 
\newcommand{\cT}{{\mathcal T}}
\newcommand{\C}{\mathbb C}
\newcommand{\E}{\mathcal E}
\newcommand{\qed}{\hfill $\Box$ \medskip}
\newcommand{\pd}{\partial}
\newcommand{\hmS}{\widehat \mS}
\newcommand{\B}{\mathbb B}
\newcommand{\ex}{\mathcal E} 
\newcommand{\tq}{\tilde q}
\newcommand{\tg}{\tilde g}
\title{{ Cyclic Statistics In Three Dimensions}}
\author{Sumati Surya  
\\
Theoretical Physics Institute, University of Alberta\\ Edmonton,
AB, Canada T6G 2G1 \\ ssurya@phys.ualberta.ca}

\begin{document}
\maketitle
\begin{abstract} 
\thispagestyle{empty} While 2-dimensional quantum systems are known to
exhibit non-permutation, braid group statistics, it is widely expected
that quantum statistics in 3-dimensions is solely determined by
representations of the permutation group. This expectation is false
for certain 3-dimensional systems, as was shown by the authors of
\cite{rafael,abks,paper}.  In this work we demonstrate the existence
of ``cyclic'', or $\Z_\N$, {\it non-permutation group} statistics for
a system of $\N > 2$ identical, unknotted rings embedded in $\re^3$.
We make crucial use of a theorem due to Goldsmith in conjunction with
the so called Fuchs-Rabinovitch relations for the automorphisms of the
free product group on $\N$ elements.
\end{abstract}

\section{Introduction}

Classical systems with configuration spaces having a non-trivial
fundamental group allow inequivalent quantisations, each labeled by
the unitary irreducible representations of this group.  A simple
illustration of this can be found in the sum-over-histories
quantisation of a particle on a circle wherein the set of paths with
fixed initial and final positions fall into classes labeled by the
winding number $m$ \cite{laidlawdewitt}.  The full partition function
is expressed as a sum of partitions over these different classes of
paths, each multiplied by an overall phase $e^{i m\theta},$ where
$\theta \in [0, 2 \pi]$ labels the unitary irreducible representations
of the fundamental group of the circle.  Each choice of $\theta$ thus
leads to an inequivalent quantisation of the system. This method of
quantisation yields several interesting phenomena ranging from the
quantum statistics of point particles, to a Hamiltonian interpretation
of the QCD theta angle, to spinorial states in quantum gravity
\cite{rafael,everyone,balbook,anyons}.

The particular phenomenon of interest to us in this paper is the
emergence of quantum statistics in systems of $\N$ identical objects
like point particles or topological geons.  The fundamental group 
$\pi_1(\Q_\N)$ of the configuration space $\Q_\N$ of such a system
contains a subgroup which, in spatial dimension $d >2 $, is isomorphic
to the permutation group on $\N$ elements $S_\N$.  On quantisation 
the unitary irreducible representations of $S_\N$ play a role in
determining the quantum statistics of the system. For $\N=2$, $d>2$,
for example, the permutation group $S_2$, generated by the exchange
operation $\E$, has two inequivalent unitary irreducible
representations: the trivial one ($\E\rightarrow 1$) corresponding to
bose statistics and the non-trivial one ($\E\rightarrow -1$)
corresponding to fermi statistics. For $\N>2$, $d>2$, $S_\N$ has
non-abelian unitary irreducible representations which give rise to
parastatistics. In dimension $d=2$, however, statistics is dictated by
an infinite discrete group, the braid group $B_\N$, rather than the
finite group $S_\N$. The resulting statistics is referred to as
``anyonic'' and plays a central role in the study of 2 dimensional
systems \cite{anyons}.

Since the permutation group $S_\N$ is always a subset of $\pi_1(Q_\N)$
for $d>2$ it is generally believed that the occurrence of
non-permutation group quantum statistics is restricted to
2-dimensions. However this is not always the case  in 3-dimensions
\cite{rafael,abks,paper}. While $S_\N$ plays a determining role in the
quantum statistics in $d>2$, it does not play the only role. As 
demonstrated in \cite{paper}, quantum statistics depends on how the
subgroup $S_\N$ ``sits'' in the the larger group $\pi_1(\Q_\N)$;
typically, $\pi_1(\Q_\N) = P \sd S_\N$, where $\sd$ denotes a
semi-direct product and $P$ is a normal subgroup of
$\pi_1(\Q_\N)$. Quantum statistics is then determined not by unitary
irreducible representations of $S_\N$, but rather those of the little
groups (or stability subgroups) $\R\subseteq S_\N$ with respect to the
action of $S_\N$ on the space of representations of $P$.

For a large class of systems the little groups are themselves
permutation subgroups $S_\M$ of $S_\N$, with $\M\leq \N$.  For
example, consider a system of 3 identical extended solitons which are
allowed to possess spin, i.e., a $2\pi$ rotation of the soliton is
non-trivial (see \cite{dyons} for an example).  Even though they are
classically identical, one can construct a representation
$\{1/2,1/2,0\}$ in which two of the solitons are spin half and the
third one is spin zero, thus rendering it quantum mechanically
distinguishable from the others. Indeed, as expected, the associated
little group of $S_3$ can be shown to be $S_2 \subset S_3$ which
corresponds to 2 rather than 3 particle quantum statistics.

However, there exist systems in which the little group need not always
be a permutation subgroup. Consider a system of $\N$ closed, identical
unknotted rings embedded in $\re^3$. Such a system could model a
collection of ring-like solitons which make their appearance in
certain non-linear sigma models \cite{hopfions}.  A crucial analogy
between this system and that of $\N$ $\re P^3$ geons in $3+1$
canonical quantum gravity was made by the authors of \cite{abks}
\footnote{Quantisation of the system of $\N$ rings has also been
examined by the authors of \cite{hl}.}. Drawing on earlier results of
\cite{rafael}, they demonstrated the existence of sectors with
indeterminate statistics for $\N =2$ \footnote{A reanalysis of these
sectors in the case of 2 $\re P^3$ geons in \cite{paper} showed that
this ambiguity originates from the lack of a canonical exchange
operator.}. The sectors we describe here are distinct in that they do
exhibit a definite, albeit non-permutation group statistics. As in
\cite{abks}, the discovery of these sectors was motivated by the
analogy with the system of $\N$ $\re P^3$ geons.  A rigorous analysis
of the quantum sectors for a system of $\N$ topological geons in $3+1$
canonical quantum gravity was carried out in \cite{paper} and the
existence of sectors obeying cyclic, or $\Z_\N$ statistics was
demonstrated for a system of $\N$ $\re P^3$ geons.  In this work we use
techniques developed in \cite{paper} to demonstrate the existence of
similar cyclic statistics\footnote{An analogue of cyclic statistics in
5-dimensions has been constructed in \cite{mick}.} for the set of $\N
\geq 3 $ closed rings embedded in $\re^3$. Namely, we show the
existence of quantum sectors in which the little group $\R$ is the
non-permutation subgroup $\Z_\N \subset S_\N$, for $\N
\geq 3$.

The inequivalent quantisations for this system of rings are determined
by the unitary irreducible representations of the so-called {\sl motion
group} $\mG$ which we present in Section 2. Using a theorem due to
Goldsmith \cite{goldsmith}, combined with the so called
Fuchs-Rabinovitch relations for the automorphisms of the free product
group on $\N$ elements \cite{FR}, we show that $\mG$ has a nested
semi-direct product structure. In Section 3 we examine the structure
of the unitary irreducible representations of a nested semi-direct
product group using Mackey's theory of induced representations
\cite{mackey}, and present our main result. We end with
some remarks in  Section 4.

Since the spin of the rings we consider is trivial, the sectors
obeying cyclic statistics clearly violate the spin-statistics
connection. In \cite{spinstats} a spin-statistics correlation was
shown to hold when the configuration space is expanded to allow the
creation and annihilation of rings, thus excluding non-permutation
group statistics. However, first quantised systems with ring-like
structures could very well occur in condensed matter systems; as
suggested in \cite{abks}, the rings can be stabilised against creation
and annihilation by carrying conserved charges. Whether sectors
obeying cyclic statistics are physically realised or not is, of course,
ultimately a question for experiment to decide.

\section{The Motion Group for a System of $\N$ Rings} 

We consider the system of $\N$ identical, non-intersecting, infinitely
thin, unknotted, unlinked, unoriented rings, $C=C_1 \cup C_2 \ldots
C_\N$ in $\re^3$, which cannot be destroyed or created.  The
configuration space $\Q_\N$ for this system of rings is the space of
embeddings of $C$ in $\re^3$ quotiented by an appropriate group of
symmetries called the {\sl motion group} $\mG$ which we will define
below.
An obvious example of a symmetry is the exchange of a pair of
identical rings. The fundamental group of $\Q_\N$ for this system is
isomorphic to the motion group $\mG$. This group is non-trivial for
all $\N\geq 1$, and has been extensively studied by Dahm and Goldsmith
\cite{goldsmith}.

Since the configuration space $\Q_\N$ is multiply connected, on
quantisation, the Hilbert space splits into inequivalent quantum
sectors.  A systematic study of such quantum sectors can be found in
\cite{balbook}.  The wavefunctions $\psi: \tQ_\N \rightarrow \C$,
where $\tQ_\N$ is the universal cover of $\Q_\N$, so that
$\pi_1(\Q_\N)$ acts non-trivially on $\psi$.  Since physically
measurable quantities like inner products should only be functions on
the classical configuration space $\Q_\N$, the action of 
$\pi_1(\Q_\N)$ on $\psi$ must be represented as a ``phase'',
which can be non-abelian for $\N\geq 2$. Thus, at every point $\tq \in
\tQ_\N$, $\psi(\tq)$ is valued in the carrier spaces of the unitary
irreducible representations of $\pi_1(\Q_\N)$. The inequivalent
unitary irreducible representations of $\pi_1(\Q_\N)$ then correspond
to inequivalent quantum sectors.

The motion group $\mG$ for this system of rings is defined as follows
\cite{goldsmith}.  Let $H(\re^3)$ denote the space of continuous maps
or homeomorphisms of $\re^3$ into itself and $H(\re^3,C)$ the subspace
of homeomorphisms which leave $C$ invariant. Let $H_\infty(\re^3)$ and
$H_\infty(\re^3,C)$ be subspaces of $H(\re^3)$ and $H(\re^3,C)$,
respectively, consisting of homeomorphisms with compact support.  A
{\it motion} is then defined as a path $h_t$ in $H_\infty(\re^3)$ such
that $h_0$ is the identity map from $\re^3$ to itself and
$h_1=H_\infty(\re^3,C)$. The product of two motions can then be
defined and the inverse $g^{-1}$ of the motion $g$ is the path
$g_{(1-t)}\circ g_1^{-1}$ \cite{goldsmith}. Two motions $h,h'$ are
taken to be equivalent if $h'^{-1}h$ is homotopic to a path which lies
entirely in $H_\infty(\re^3,C)$.  The motion group $\mG$ is then the
set of equivalence classes of motions of $C$ in $\re^3$ with
multiplication induced by $\circ$\footnote{For brevity of expression
we will henceforth refer to an equivalence class of motions as a
motion.}.

We will use Hendricks' definition of a rotation \cite{hendricks} to
describe the generators of the motion group. A 3-ball $\B^3 \subset
\re^3$ will be said to be rotated by an angle $\alpha$ in the
following sense: take a collar neighbourhood $S^2\times [0,1]$ of $\pd
\, \B^3 \approx S^2$ and let the $S^2$'s be differentially rotated
from $0$ to $\alpha$ with $S^2\times \{0\}=\pd \, \B^3$ rotated by
$\alpha$ and $S^2\times \{1\}$ not rotated at all. The rotation by an
angle $\alpha$ of a solid torus $U=\B^2 \times S^1$ in the direction
of its non-contractible circle $S^1$ is similarly defined as a
differential rotation of a collar neighbourhood $\,T^2 \times [0,1]$ of
$\pd U\approx T^2$, with $T^2\times \{0\}=\pd U$ rotated by $\alpha$
and $T^2\times \{1\}$ not rotated at all.

$\mG$ is generated by three types of motions which are quite easily
visualised \cite{goldsmith}.  The first is the flip motion $f_i$ which
corresponds to ``flipping'' the $i$th ring \footnote{In the case of
oriented rings, this motion yields a configuration distinct from the
first and is not a symmetry.}.  This motion corresponds to a rotation
by $\pi$ of an open ball in $\re^3$ containing $C_i$, about an axis
lying in the plane of $C_i$. Since the rings are embedded in three
dimensions, $f_i^2=e$, so that each flip generates a $\Z_2$ subgroup.
Next is the exchange motion $e_i$ which exchanges the $i$th ring with
the $(i+1)$th ring. This can be thought of as a $\pi$ rotation of a
solid torus in $\re^3$ containing both $C_i$ and $C_{i+1}$ (but no
others).  These motions generate the permutation group
$S_\N$. Finally, one has the slide motion $s_{ij}$ which requires a
slightly more detailed description. A point in the configuration space
(i.e. $\re^3 -C$ modulo the action of the motion group) is itself a
multiply connected space with $\pi_1(\re^3-C)$ isomorphic to the free
product group on $\N$ generators
$F(x_1,x_2 \ldots, x_\N) \approx \Z*\Z\ldots *\Z$, each factor of $\Z$
isomorphic to the fundamental group of a single ring in
$\re^3$. $s_{ij}$ is then the motion of $C_i$ along one of
these $\Z$ factors, specifically, the generator of $\Z\subset
\pi_1(\re^3-C)$ passing through $C_j$. Again, one can define the
slide using a rotation: consider a solid torus containing $C_i$ and
``threading'' $C_j$, without intersecting it. A slide is then a $2\pi$
rotation of this solid torus.  The existence of slide motions is key
to the present analysis, and is what makes the analogy with the system
of topological geons explicit.

We denote the three subgroups generated by the flips, the exchanges
and the slides as $\mF$, $S_\N$ and $\mS$, respectively. We will also
need to identify the subgroup $\tG$ generated by only the flips and
the exchanges.  The structure of $S_\N$ is known: it is simply the
permutation group on $\N$ elements. However, the structures of $\mF$
and $\mS$ need to be deduced, as does information on how these groups
sit in $\mG$.  While the generators of $\mG$ have been known for some
years, its structure has not been obtained until now.  We now show
that $\mG$ has the nested semi-direct product structure
\begin{equation}
\mG = \mS \sd (\mF \sd S_\N). \label{sdproduct} 
\end{equation} 
We also show that $\mS$ is the non-abelian group made up of the free product
group on $\N(\N-1)$ generators
\begin{equation}
\label{freeprod.eq}
\underbrace{\Z * \Z * \ldots *\Z}_{\N(\N-1)},   
\end{equation} 
subject to the conditions 
\begin{equation} 
\label{slide.eq}
s_{ij}s_{kl}=s_{kl}s_{ij}, \quad
s_{ij}s_{kj}=s_{kj}s_{ij}, \quad
s_{ik}s_{jk}s_{ij}=s_{ij}s_{ik}s_{jk}.   
\end{equation} 
$\mF$, on the other hand, can be shown to be the abelian group
isomorphic to the direct product group of the $\Z_2$ flips of each
ring
\begin{equation}
\label{flips.eq}
\mF=\underbrace{\Z_2 \times \Z_2\times  \ldots \times \Z_2}_\N.  
\end{equation}

\noindent {\bf Lemma:} {\sl $\mG$ has the nested semi-direct product
structure (\ref{sdproduct}). Moreover, $\mS$ is the group
(\ref{freeprod.eq}) subject to the conditions (\ref{slide.eq}), and
$\mF$ is the group (\ref{flips.eq}).}

\noindent {\bf Proof:} 
The induced action of the motion group on $\pi_1(\re^3-C)$ has been
examined by Goldsmith \cite{goldsmith}, and provides us with a
crucial step in deducing the structure of $\mG$. As noted
earlier, $\pi_1(\re^3-C)$ is isomorphic to $ F(x_1, \ldots, x_\N)$, the free
product group on $\N$-generators, $x_i,\,\, i= 1, \cdots \N$. 
In \cite{goldsmith} the ``Dahm''  homomorphism $D: \mG \rightarrow
Aut(F(x_1, \ldots, x_\N))$ is defined where $Aut(F(x_1, \ldots, x_\N))$
is the  group of automorphisms of $F(x_1, \ldots, x_\N)$. For each
motion $g\in \mG$, $D$  induces an automorphism of $F(x_1,
\ldots, x_\N)$.  The following theorem then states: 

\noindent {\bf Goldsmith's Theorem \cite{goldsmith}:} {\sl The group of
motions $\mG$ of the trivial $\N$-component link $C$ in
$\re^3$ is generated by the following types of motions:
\begin{enumerate}
\item $f_i$ or flips. Turn the $i^{th}$ ring over. This induces the
automorphism $\phi_i: x_i \rightarrow x_i^{-1}, \, \, x_k \rightarrow
x_k$, $ k\neq i$,  of $F(x_1, \ldots, x_\N)$.
\item $e_i$ or exchange. Interchange the $i^{th}$ and the $(i+1)^{th}$
rings. The induced automorphism of $F(x_1, \ldots, x_\N)$ is
$\epsilon_i: x_i \rightarrow x_{i+1}, x_{i+1} \rightarrow x_i$ and
$x_k =x_k$ for $k \neq i,i+1$.
\item $s_{ij}$ or slides. Pull the $i^{th}$ ring through the $j^{th}$
ring. This induces the automorphism $\sigma_{ij}:x_i \rightarrow
x_jx_ix_j^{-1},$  $x_k \rightarrow x_k$,  $k \neq i$,  of
$F(x_1, \ldots, x_\N)$.
\end{enumerate}
Moreover, the Dahm homomorphism, $D:\mG \rightarrow Aut(F(x_1, \ldots,
x_\N))$ is an isomorphism onto the subgroup $\oG$ of $Aut(F(x_1\ldots
x_\N))$ generated by $\phi_i, \epsilon_i$ and $ \sigma_{ij}$, where
$1\leq i,j\leq n, i\neq j$. } 

Let us denote the subgroups of $\oG$ generated by the automorphisms
$\sigma_{ij}$, $\phi_i$ and $\epsilon_i$ as $\oS$, $\oF$ and $\oSn$,
respectively.  We may now employ the Fuchs-Rabinovitch relations for
$Aut(F(x_1, \dots , x_n))$ which provides a complete set of relations
for the generators of $\oG$ \cite{FR}.  For $\pi_1(\re^3-C) = \Z * \Z
\ldots * \Z$, in particular, these relations are simple and imply that
$\oG \subset Aut(F(x_1, \dots , x_n))$ has the nested semi-direct
product structure
\begin{equation}
\label{gbar.eq} 
\oG=\oS \sd(\oF \sd \oSn) = \oS \sd \otG, 
\end{equation}  
where $\otG= \oF \sd \oSn$. Moreover, these relations imply that $\oS$
is the free product group on $\N(\N-1)$ generators $\Z * \Z * \ldots
*\Z$ subject to the conditions
$\sigma_{ij}\sigma_{kl}=\sigma_{kl}\sigma_{ij},$
$\sigma_{ij}\sigma_{kj}=\sigma_{kj}\sigma_{ij},$ and
$\sigma_{ik}\sigma_{jk}\sigma_{ij}=\sigma_{ij}\sigma_{ik}\sigma_{jk}$
while $\oF$ is the abelian direct product group made up of $\N$
factors of $\Z_2$, $\oF=\Z_2 \times \Z_2\times \ldots \times \Z_2$.
Since $D$ is an isomorphism with $D(\mS)\subseteq \oS$,
$D(\mF)\subseteq \oF$ and $D(S_\N)\subseteq \oSn$, this means that
$\mS \approx \oS$, $\mF \approx \oF$ and $S_\N \approx \oSn$.  From
(\ref{gbar.eq}), it is then obvious that $\mG$ itself has the nested
semi-direct product structure (\ref{sdproduct}).  Moreover, $\mS$ is
the free product group on $\N(\N-1)$ generators $\Z * \Z *
\ldots *\Z$ subject to the relations (\ref{slide.eq}) and $\mF$ is
given by (\ref{flips.eq}).  \qed

While the structure of the motion group can be completely deduced from
the Dahm homomorphism and the Fuchs Rabinovitch relations, it is
instructive to examine this group without recourse to $Aut(F(x_1,
\dots , x_n))$. Using just the definition of the motion group we now
illustrate the following properties of $\mG$: (a) $\mS$ is normal in
$\mG$ and satisfies the relations (\ref{slide.eq}) and (b) that $\mF$
is normal in $\tG$.

By definition, an element of the motion group is a homotopy
equivalence class of paths in the space of homeomorphisms with compact
support. Two homeomorphisms $h_1$ and $h_2$ with compact support on
the regions $U_1$ and $U_2$ commute if $U_1 \cap U_2 =\phi$ and hence
so do the corresponding motions. It is therefore useful to isolate the
``minimal'' neighbourhoods in which homeomorphisms representing the
generators of the motion group act so as to determine which two
motions commute.

Let $U_i$ denote an open ball  neighbourhood of $C_i$ in $\re^3$
which contains no other $C_j$, $j\neq i$, and let $U_{ij}$ denote an
open ball neighbourhood of $C_i \cup C_j$ containing no other $C_k$,
$k\neq i,j$, etc. We will refer to the $U_i$ as ``exclusive''
neighbourhoods and the $U_{ij}, U_{ijk}, \ldots $ etc.  as ``common''
neighbourhoods. The flip motion $f_i$ is then defined by a homotopy
equivalence class of paths in $H_\infty(\re^3)$ which include a
``model'' path made up of homeomorphisms with support only on $U_i$,
i.e., a path in $H_\infty(\re^3)$ along which $C_i$ is flipped without
disturbing any of the other rings. Next, the exchange motion $e_i$ is
defined by a homotopy equivalence class of paths including a model path
made up of homeomorphisms with support only on $U_{i(i+1)}$, i.e., the
$i$th and the $(i+1)$th ring are exchanged without disturbing the
other rings.  Finally, the slides $s_{ij}$ are defined by a homotopy
equivalence class of paths including a model path with support only on
$U_{ij}$, i.e., a path in which the other rings are not disturbed.

Now, the set of exclusive neighbourhoods $\{U_i\}$ remains invariant
when acted upon by the subgroup $\tG$ generated by the flips and by
the exchanges. This is obvious for $\mF$, since each flip $f_i$ acts
within an exclusive neighbourhood. For $S_\N$, while the exchange
$e_i$ has compact support on $U_{i(i+1)}$, its action can be
considered as a pure exchange of $U_i$ with $U_{i+1}$. Thus, one can
consider as a model path for the exchange, a localised $\pi$ rotation
in $U_{i(i+1)}$ which exchanges $U_i$ with $U_{i+1}$.  This, however,
is not the case with the slides $s_{ij}$. While the set $\{U_k \}$ for
$k\neq j$ remains invariant under the slide $s_{ij}$ of $C_i$ through
$C_j$, the exclusive neighbourhood $U_j$ does not. The non-local
action of the slide takes $U_j$ into a set $V_j$ which ``encloses''
$C_i$ even though it does not contain it, i.e. there exists a $U_i$
such that $U_i\cap V_j =\phi$ (see fig). Thus, $V_j$ is not an
exclusive neighbourhood of $C_j$. This feature leads to subtleties in
what follows. 
\begin{figure}[ht]
\centering
\resizebox{2.0in}{!}{\includegraphics{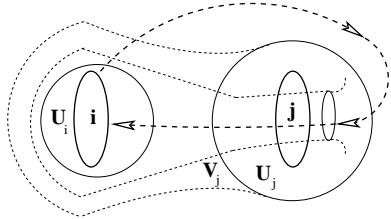}}
\vspace{0.5cm}
\caption{Under the slide $s_{ij}$ $C_i$ ``tunnels'' through
  the neighbourhood $U_j$ of $C_j$ and maps it onto the region $V_j$,
  shown by dashed lines. $V_j$ therefore ``encloses'' $C_i$ without
  containing it, i.e. $U_i\cap V_j=\phi$. }
\end{figure}

Since the exchanges and the flips leave the set $\{U_i\}$ invariant,
model paths are sufficient to see that $\mF$ is normal in $\tG$, 
i.e.,  for all $\tg\in \tG$, $i \leq \N$, $\,\tg f_i \tg^{-1} \in \mF$.
To show this, it is sufficient to take $\tg$ to be an exchange.  For
the motion $e_i f_j e_i^{-1}$ with $j \neq i,i+1 $, the model paths
for $e_i$ and $f_j$ have compact support on $U_{i(i+1)}$ and $U_j$,
respectively, where $U_{i(i+1)}\cap U_j =\phi$. Hence the motions
commute, so that $e_i f_j e_i^{-1}=f_j$. Now consider the motion $e_i
f_i e_i^{-1}$. The model paths for $e_i^{-1}$ exchange $U_i$ with
$U_{i+1}$. One can then use a model path for the motion $f_i$ which
acts on some $U_{i+1}' \subset U_{i+1}$ so that the final exchange
$e_i$ which exchanges $U_{i+1}$ with $U_i$ does not disturb the action
of $f_i$ on $U_{i+1}'$. Thus, $e_i f_i e_i^{-1}=f_{i+1}$. Similarly,
$e_i f_{i+1} e_i^{-1}=f_{i}$.

However, model paths are insufficient when one wants to deal with the
slides. Let us consider a motion whose model path involves
homeomorphisms with support only on the compact region $U$.  The
homotopy class of paths defining this motion also includes the
non-model, or ``gregarious'' paths, which involve homeomorphisms with
non-trivial compact support on $\re^3-U$. In other words, gregarious
paths {\it can} disturb the other rings; they can contain
homeomorphisms with non-trivial support on neighbourhoods of rings
left undisturbed by the model path.  Consider the motion $f_i$ for
simplicity. A model path for $f_i$ has support only on $U_i$ and
corresponds to a $\pi$ rotation about an axis in the plane of $C_i$. A
gregarious path on the other hand can be constructed piecewise as
follows: (a) rotate $C_i$ by $\pi/3$, about an axis $\hat x$ in its
plane (b) flip another $C_j$, $j\neq i$, (c) rotate $C_i$ by a further
$\pi/3$ about $\hat x$ in the same sense as before (d) flip $C_j$
again, (e) and complete with a further $\pi/3$ rotation of $C_i$ about
$\hat x$ in the same sense as before.  Such a path clearly corresponds
to the motion $f_i$, but involves homeomorphisms of $\re^3$ in
$H_\infty(\re^3)$ which have non-trivial support on the ring $C_j$,
$j\neq i$.

Both model and gregarious paths are necessary to demonstrate that
$\mS$ is normal in $\mG$.  $\mS$ is a normal subgroup of $\mG$ if
$\forall g \in \mG$ and $\forall i, j \leq n$ $g s_{ij} g^{-1} \in
\mS$. It is sufficient to take $g$ to be a generator of $\rSn$ or
$\mF$. 

We begin with the exchanges. Let us examine the motion $e_k s_{ij}
e_k^{-1}$ by considering only model paths in the appropriate homotopy
class. For $k\neq i,j$ $e_k s_{ij} e_{k}^{-1}=s_{ij}$, since the
homeomorphisms that make up model paths for $e_k$ and $s_{ij}$ have
compact supports on $U_k$ and $U_{ij}$ with $U_k \cap U_{ij}=\phi$.
Model paths are, however, insufficient to show that
$e_ks_{ij}e_k^{-1}$ is also a slide for $k=i$, $i-1,$ $j$ or
$j-1$. Consider the motion $e_j s_{ij} e_{j}^{-1}$ with $k=j$, $i \neq
j, j+1$ $e_{j}^{-1}$ swaps $U_j$ with $U_{j+1}$ by a $\pi$ rotation of
a torus containing both $U_j$ and $U_{j+1}$. Next, $s_{ij}$ rotates by
$2\pi$ a solid torus containing $C_i$ and threading $C_{j+1}$, thus
mapping $U_{j+1}$ into a non-exclusive neighbourhood $V_{j+1}$. A
model path for the final exchange $e_j$ would rotate by $\pi$ a solid
torus containing new exclusive neighbourhood $U'_{j+1}$ of $C_{j+1}$
and $U_i$. $U_{i(j+1)}$ in which the slide acts, is not left invariant
by this final exchange, making the resultant motion difficult to
unravel.  Instead, we use the following gregarious path to perform the
final exchange: consider a path in $H_\infty(\re^3)$ where $U_{i(j+1)}$
and $U_j$ are swapped by performing an appropriate $\pi$ rotation in
the common neighbourhood $U_{ij(j+1)}$ of $C_i, C_j$ and $C_{j+1}$.
The final exchange motion is then completed by merely moving $C_i$
back to its original position. $U_{i(j+1)}$ is thus left undisturbed
so that the full motion is the slide $s_{i(j+1)}$.  A use of a similar
gregarious path for the final exchange shows that $e_{(j-1)} s_{ij}
e_{(j-1)}^{-1} = s_{i(j-1)}$, $e_{i} s_{ij} e_{i}^{-1} = s_{(i+1)j}$
and $e_{(i-1)} s_{ij} e_{(i-1)}^{-1} = s_{(i-1)j}$.

Next, consider the flips. The motion $f_k s_{ij} f_k^{-1}$ can again
be examined using only model paths for $k\neq j$, and we can see that
it is $s_{ij}$. This is because the model path for $f_k$ has compact
support only on $U_k$ which is undisturbed by the slide even when
$k=i$.  However, the use of model paths is insufficient to examine the
motion $f_j s_{ij} f_j^{-1}$: not only does $U_j$ not remain an
exclusive neighbourhood under the slide $s_{ij}$, but the $f_j$ moves
the points in $U_j$ relative to each other. Rather than consider just
a single gregarious path, following \cite{paper}, we use a particular
set of homotopy equivalent paths. Let $\kappa$ be the generator of
$\pi_1(\re^3-C)$ through $C_j$ about which the slide $s_{ij}$ takes $C_i$.
We define the paths $\gamma_\alpha$ as follows: (a) perform a ``part''
inverse flip corresponding to a $(\pi-\alpha)$ rotation of $C_j$ about
$\hat x$ (b) slide  $C_i$ through $C_{j}$ along
$\kappa^{-1}$ (c) finish the inverse flip $f_j^{-1}$ of $C_j$ by a
rotation $\alpha$ about $\hat x$ and (d) finally, perform the flip
$f_j$ of $C_j$ about $\hat x$.  $\gamma_0$ then corresponds to the
model path for the motion $f_j s_{ij}f_j^{-1}$ while the path
$\gamma_{\pi}$ corresponds to the slide $s_{ij}^{-1}$. Since $\alpha$
is a continuous parameter $\alpha \in [0,\pi]$, the $\gamma_\alpha$
provide a homotopy map from $\gamma_0$ to $\gamma_{\pi}$, which
implies that $f_j s_{ij}f_j^{-1} = s_{ij}^{-1}$ 
\footnote{It is perhaps a useful exercise for the reader to see why a
  similar argument cannot be used to find a set of homotopic paths
  between $s_{ij} f_j s_{ij}^{-1}$ and an element of $\mF$.}.

Thus, the slide subgroup  $\mS$ is a normal subgroup of $\mG$.

We can also demonstrate that the relations (\ref{slide.eq}) are
satisfied by $\mS$, using just the definition of the motion group.
The first of these relations is clearly satisfied by the generators of
$\mS$, since the model paths corresponding to the slides $s_{ij}$ and
$s_{kl}$ involve homeomorphisms with compact support only on $U_{ij}$
and $U_{kl}$ where $U_{ij} \cap U_{kl} =\phi$. It takes a little more
work to show that the other two relations are also satisfied by the
generators of $\mS$.

Consider the motion $s_{ij}s_{kj}s_{ij}^{-1}$. $s_{ij}$ and $s_{kj}$
are slides of the two rings $C_i$ and $C_k$ through a third ring
$C_j$. These slides are obtained by $2\pi$ rotations of the solid
tori $V_{ij} \approx B^2 \times S^1$ and $V_{kj} \approx B^2 \times
S^1$ which thread through $C_j$, with $V_{ij} \cap V_{kj} = \phi$.
Define the paths $\gamma_\alpha$ as follows: (a) a rotation by
$-\alpha$ of $V_{ij}$ (b) a $2\pi$ rotation of $V_{kj}$ (c) a $-(2\pi
-\alpha)$ rotation of $V_{ij}$ and finally, (d) a $2\pi$ rotation of
$V_{ij}$. $\gamma_0$ then defines a model path for the motion
$s_{ij}s_{kj}s_{ij}^{-1}$, and $\gamma_{2\pi}$ corresponds to the
slide $s_{kj}$.  Since $\alpha$ is a continuous parameter, $\gamma_0$
is homotopic to $\gamma_{2\pi}$ and hence also corresponds to
$s_{kj}$. Notice that by keeping $V_{ij} \cap V_{kj}=\phi$ we prevent
a mixing of their rotations and hence the deformations of the
neighbourhood $U_j$ by $s_{ij}$ and by $s_{kj}$.
 
Next, consider the motion $s_{ij}s_{ik}s_{jk}s_{ij}^{-1}$. Although
this looks considerably more complicated than the previous motion, the
two elements of $\mS$ involved, $s_{ik}s_{jk}$ and $s_{ij}$, have
compact supports on non-intersecting neighbourhoods. Namely, the
element $s_{ik}s_{jk}$ corresponds to sliding $C_j$ through a
generator $\rho$ of $\pi_1$ of $C_k$ and then sliding $C_i$ through
the same generator. Under this action, $U_j \rightarrow U_j$ and $U_i
\rightarrow U_i$, while $U_k$ is now mapped to a region $V_k$ which
now ``encloses'' both $C_i$ and $C_j$. Thus, there exists a path in
$H_\infty(\re^3)$ corresponding to the motion $s_{ik}s_{jk}$ made up
of homeomorphisms which leave the common neighbourhood $U_{ij}$
undisturbed. Since there is a model path corresponding to the slide
$s_{ij}$ which has compact support only on $U_{ij}$, this means that
the two motions $s_{ik}s_{jk}$ and $s_{ij}$ indeed commute.  Thus, the
generators of $\mS$ satisfy all the relations (\ref{slide.eq}).

\noindent {\bf Remark:} In \cite{abks} a set of relations for the
generators in the $n=2$ case was given:
$f_i^2=\ex^2=(f_i\ex)^4=(f_i\ex s_j \ex)^2=e $ where $i=1,2$ and the
slides $s_i$ generate $\mS$, the flips $f_i$ generate $\mF$ and the
exchange $\ex$ generates $S_2$.  These follow in a straightforward
manner from the relations presented above. 

\section{Cyclic Statistics} 

The inequivalent quantum sectors for our system of $\N$ identical
rings are labeled by the unitary irreducible representations of
$\pi_1(\Q_\N) \approx \mG$.  The group $\mG$ represents a gauge
symmetry and the action of the individual motions $g\in \mG$ on
$\re^3-C$ can be used to interpret the associated quantum phases.  For
example, for a single ring the motion group is simply $\mF=\Z_2$,
which has two unitary irreducible representations: the trivial one and
the non-trivial one. The associated quantum theories thus correspond
to an ``unoriented'' quantum ring in which wavefunctions $\psi$
transform as $\psi \rightarrow \psi$ under a flip, and an ``oriented''
quantum ring in which $\psi \rightarrow -\psi$ under a
flip. Similarly, for $\N \geq 2$ the motions corresponding to
permutations of the rings can be non-trivially represented, and lead
to different quantum statistics (see \cite{rafael,paper,balbook} for a
more detailed discussion of quantum phases and statistics for
extended objects).

As mentioned in the introduction, the quantum statistics of a system
is not solely determined by $S_\N$, but rather by the unitary
irreducible representations of its stability subgroup $R\subseteq S_N$
associated with its action on the unitary irreducible representations
of the normal subgroup $\mS \sd \mF$ of $\mG$. This follows from
Mackey's theory of induced representations for semidirect product
groups $P \sd K$ \cite{mackey}.  In this construction, one begins with
the space of unitary irreducible representations $\hN$ of the normal
subgroup $P$.  The subgroup $K$ has the (not necessarily free) action
on $\hN$
\begin{equation} 
\Delta(p) \rightarrow \widetilde \Delta(p)= \Delta(kpk^{-1}), 
\end{equation} 
where $\Delta \in \hN$, $p\in P$ and $k\in K$. Starting with a
particular $\Delta_1 \in \hN$ one obtains an orbit ${\mathcal O}=\{
\Delta_1, \Delta_2, \ldots \Delta_r\}$ of the $K$ action on $\hN$, and
the little group $\R$ associated with $\mathcal O$.  The full unitary
irreducible representation of $P\sd K$ is then built up by taking the
direct product of (a) $[ \Delta_1 \oplus \Delta_2 \oplus\ldots\oplus
\Delta_r] $ with (b) a unitary irreducible representation of $\R$.
For example, if one starts with the trivial representation of $P$,
then the orbit consists of a single point and $\R=K$. The unitary
irreducible representation of $P \sd K$ that can be constructed from
this orbit are just the unitary irreducible representations of $K$. On
the other hand, one may find an orbit of $K$ in $\hN$ with $\R=e$. The
full unitary irreducible representation is then simply the sum of the
unitary irreducible representations in the orbit, $\oplus_i
\Delta_i$. The action of the subgroup $K$ is then reduced to a
canonical map which permutes the carrier spaces $\Hil_i$ of $\Delta_i$
\cite{paper} 
\footnote{ As discussed in \cite{paper} 
for $\N \geq 4$  the possibility of projective statistics exists when
$\pi_1(Q)$ has a semi-direct product structure.}.

We now illustrate the importance of the little group in determining
quantum statistics with a simple example.  Because of the nested
semi-direct product structure of the motion group, we may begin by
first representing the slides trivially.  We thus need to find only
the unitary irreducible representations of the subgroup $\tG = \mF \sd
S_\N$. Since $\mF \approx \Z_2 \times \Z_2 \ldots \times \Z_2$, it is
trivial to list its unitary irreducible representations, $\Delta\equiv
(k_1, k_2 \ldots k_\N)$, with $k_i =\pm 1$. For example, for $\N=3$,
let us start with the unitary irreducible representation
$\Delta_1=(-,-,+)$ of the normal subgroup $\mF$ of $\tG$. This choice
corresponds to two of the rings being identical and oriented, while
the third is unoriented and hence distinguishable from the others.
The action of $S_3$ on $\Delta_1$ generates the orbit $ \{\Delta_1,
\Delta_2,\Delta_3 \} \equiv\{(-,-,+), (+,-,-), (-,+,-)\}$ in $\hmF$
whose associated little group is $S_2$.
The resulting unitary irreducible representation of $\tG$ is then
$\{\Delta_1 \oplus \Delta_2 \oplus \Delta_3\} \otimes \Gamma$, where
$\Gamma$ is a unitary irreducible representation of $S_2$. Under a two
particle exchange $\Gamma$ provides either a bosonic (+1) or a
fermionic (-1) phase.  Since one of the three rings has been rendered
quantum mechanically distinguishable from the other two, one obtains an
appropriate two ring statistics. The action of the remaining elements of
$S_3$, namely the cyclic elements, is canonical: they merely
permute the carrier spaces $\Hil_i$ of the $\Delta_i$.  This general
structure continues to hold for all $\N$, and is illustrative for the
case of primary interest here when the slides are non-trivially
represented.

Before proceeding to construct a quantum sector exhibiting cyclic
statistics for $\N \geq 3$, let us consider the simplest case with the
slides non-trivially represented, namely when $\N =2$.  For $\N=2$,
the slide subgroup is generated by the two slides $s_1, s_2$, the flip
subgroup $\mF$ by the two flips $f_1, f_2$ and the permutation group
$S_2$ by the exchange $\ex $.  The following example demonstrates a
peculiar feature which will reappear for $\N > 2$, whereby slides
render a pair of ``locally identical'' rings distinguishable.  Let us
start with the abelian unitary irreducible representation of $\mS$,
$\hS_1(s_1)= 1,\, \hS_1(s_2)=-1$.  The action of $f_j$ on $\hS_1$ is
$\hS_1(s_i) \rightarrow \widetilde \hS_1(s_i)=\hS_1
(f_js_if_j^{-1})=\hS_1(s_i)$ and is hence contained in the little
group $\R$ of $\tG$. Under the action of $\ex $, $\hS_1(s_i)
\rightarrow \widetilde \hS_1(s_i)=\hS_1 (\ex s_i\ex^{-1})=\hS_1(s_j)
\neq \hS_1(s_i)$ where $j \neq i$, so that $S_2 \nsubseteq \R$. Thus,
the two rings are quantum mechanically distinguishable even if $\mF$
is trivially represented. This is very unusual, since
indistinguishability of a collection of objects is often thought of as
a local, intrinsic property of each object.  However, in this
representation, it is the non-local action of slides which renders the
two rings distinguishable: the rings slide through each other
differently. Thus, there exist a wavefunction $\psi$ peaked on a
configuration of two well-separated rings such that under the action
of $s_1$, $\psi \rightarrow \psi$ and under that of $s_2$, $\psi
\rightarrow -\psi$.  This quantum lifting of indistinguishability by
slides is what leads to non-permutation group statistics for $\N > 2$.

Let us begin with the case $\N=3$. $\mS$ is generated by the six
generators $s_{ij}$, $i,j=1,2,3$, $i\neq j$, $\mF$ is generated by the
$3$ elements $f_1, f_2, f_3$, and the permutations form the
non-abelian subgroup $S_3$. We start with the following abelian
unitary irreducible representation $\hS_1$ of $\mS$:
\begin{equation} 
\hS_1(s_{12}) = \hS_1(s_{23})= \hS_1(s_{31}) = -1, \quad 
\hS_1(s_{21}) = \hS_1(s_{32}) = \hS_1(s_{13})  = 1.  \label{rept.eq}
\end{equation} 
Consider the action of $\tG$ on $\hS_1$.  The action of a flip $f_k$
on $\hS_1$ for $k=i$ or $j$ is: $ \hS_1(s_{ij}) \rightarrow
\hS_1(f_ks_{ij}f_k^{-1})=\hS_1^{-1}(s_{ij})=\hS_1(s_{ij})$, while the
action of $f_k$, $k\neq i, j$ is trivial. Thus, $\mF$ lies in the
stability subgroup of $\tG$. On the other hand, the exchanges $e_i$ do
not leave $\hS_1$ invariant: $ \hS_1(s_{i(i+1)}) \rightarrow \hS_1(e_i
s_{i(i+1)}e_i^{-1})=\hS_1(s_{(i+1)i})=-\hS_1(s_{i(i+1)})$.  Curiously,
however, there are elements of $S_3$ which leave $\hS_1$ invariant,
namely the subgroup of cyclic permutations $\Z_3$ of $S_3$ generated by
$z=e_2e_3$. Under the action of $e_2e_3$ the slides $ \{s_{12},
s_{23}, s_{31}\}\rightarrow \{s_{23}, s_{31}, s_{12}\}$, and
$\{s_{21}, s_{32}, s_{13}\} \rightarrow \{s_{32}, s_{13}, s_{21}\}$,
which leaves $\hS_1$ invariant. Thus, the stability subgroup is
$\mF\sd \Z_3$.  The remaining elements of $\mF\sd S_3$ $e_{1},e_{2}$
and $ e_{3}$ generate the two element orbit $\mathcal O \equiv
\{\hS_1,\hS_2\}$ in $\hmS$ the space of unitary irreducible
representations of $\mS$, where
\begin{equation} 
\hS_2(s_{12}) =  \hS_2(s_{23})= \hS_2(s_{31}) = 1, \quad
\hS_2(s_{21}) =  \hS_2(s_{32}) = \hS_2(s_{13})  = -1. \label{orbit.eq}
\end{equation} 
The associated unitary irreducible representation  of $\mG$ is
therefore  non-abelian, and can be symbolically expressed as 
\begin{equation} 
(\hS_1 \oplus \hS_2) \otimes \cT,  
\end{equation} 
where $\cT$ is a unitary irreducible representation of the stability
subgroup $\mF\sd \Z_3$. 

Let us for simplicity consider the case when $\mF$ is trivially
represented in $\cT$, so that $\cT$ is a unitary irreducible
representation of $\Z_3$. $\Z_3$ has two non-trivial inequivalent
unitary irreducible representations (a) $z \rightarrow e^{\frac{2\pi
i}{3}}$ and (b) $z \rightarrow e^{\frac{4\pi i}{3}}$. Thus, there
exist wavefunctions $\psi_a,$ $\psi_b$ in the corresponding quantum
sectors which are peaked on a configuration of well separated rings
and which pick up the phases $\psi_a \rightarrow e^{\frac{2\pi
i}{3}}\psi_a$ and $\psi_b \rightarrow e^{\frac{4\pi i}{3}}\psi_b$,
respectively, under the action of the cyclic permutations. Thus, these
sectors exhibit a cyclic, non-permutation group, statistics: the rings
are identical {\it only} when permuted by a cyclic combination, and
{\it not} under pair-wise exchange!  This is indeed a very curious
behaviour and is, again, linked to the non-locality of slide
motions: even though the flips are all trivially represented
the slides render the rings pair-wise distinguishable but cyclically
indistinguishable. We will say that the rings obey $\Z_3$ {\it cyclic
statistics}.

The case for arbitrary $\N>2$ follows in a straightforward
manner.Namely, we can always isolate a pair of non-trivial subsets
from the set of slide generators $\{s_A\}$ and $\{s_B\}$ which are
invariant under $\Z_{\N}$.  There is a small difference in the
construction in the even $\N=2m$ and odd $\N=2m+1$ cases. For $\N=2m$,
$\Z_{2m}$ contains the subgroup $\Z_2$; if $z$ is the generator of
$\Z_{2m}$ with $z^{2m}=e$, then $z^m$ generates a $\Z_2$ subgroup
corresponding to $m$ commuting exchanges. One can then see that the
two sets of generators $\{s_A\}$ and $\{s_B\}$ which are invariant
under $\Z_{2m}$ have cardinality $2m(m-1)$ and $2m^2$
respectively. For $\N=2m+1$, $Z_2$ is not a subgroup of
$\Z_{2m+1}$. Hence the two sets of generators $\{s_A\}$ and $\{s_B\}$
each have cardinality $m(2m+1)$. One can thus obtain $\Z_\N$ cyclic
statistics for arbitrary $\N>2$.

We end this section by commenting on the possibility that sectors with
more complicated non-permutation group statistics may exist. To construct
the above cyclic statistics sectors we started with a very simple
abelian unitary irreducible representations of the slide subgroup.  It
is conceivable that if one instead started with a non-abelian unitary
irreducible representation of $\mS$ (with certain symmetries) that the
stability subgroup $\mF \sd K$ associated with it 
is such that $K$ is non-abelian and a non-permutation 
subgroup of $S_\N$. Such a sector would then exhibit a {\it
non-abelian, non-permutation} group statistics.  Our current work
provides a framework in which to probe such questions.

\section{Remarks} 

Anyonic statistics in $2+1$ dimensions can be
modeled by adding a Chern Simon's term to the $\N$ particle Lagrangian
\cite{cs}.  In \cite{abks} a stringy generalisation of this was
developed to obtain non-trivial phases from the action of the motion
group, namely a $B\wedge F$ topological term made up of an abelian
gauge field and an axion field was added to the $\N$ string Lagrangian
along with an interaction term. Similar systems have subsequently been
studied in \cite{hl}. In \cite{abks} it was shown that even though the
statistical phases are trivial (i.e. bosonic) the action of the slide
subgroup is non-trivial, giving rise to fractional quantum
phases. Since slides involve the motion of one ring through a
non-trivial generator of the fundamental group of another ring, these
fractional phases correspond to Aharnov-Bohm phases rather than to
fractional quantum statistics. Indeed, slides can occur between
non-identical particles as well and hence the interpretation of such
phases as statistics in \cite{hl} seems questionable. Since cyclic
statistics occur in non-abelian sectors of the system, it would be
interesting to construct appropriate non-abelian generalisations of
\cite{abks} which exhibit this behaviour. We leave this problem to
future investigations.


\section{Acknowledgments} 
I would like to thank A.P. Balachandran and  Rafael Sorkin  for
discussions  
and for valuable comments on an earlier draft. I would also like to
thank Ajit Srivastava for discussions during the early stages of this work
and
the Physics Department at U.C. Davis for its hospitality while this
work was being written up. This work has been supported in part by a
grant from the Natural Sciences and Engineering Research Council
(Canada).

\end{document}